# Modified quantum-speed-limit bounds for open quantum dynamics in quantum channels

Xin Liu,[1] Wei Wu,[1,*] and Chao Wang[2]

[1]*Department of Physics, School of Science, Wuhan University of Technology（WUT）, Wuhan, China*

[2]*School of Engineering and Digital Arts, University of Kent, Canterbury, United Kingdom*

The minimal evolution time between two distinguishable states is of fundamental interest in quantum physics. Very recently Mirkin *et al.* argue that some most common quantum-speed-limit (QSL) bounds which depend on the actual evolution time do not cleave to the essence of the QSL theory as they grow indefinitely but the final state is reached at a finite time in a damped Jaynes-Cummings (JC) model. In this paper, we thoroughly study this puzzling phenomenon. We find the inconsistent estimates will happen if and only if the limit of resolution of a calculation program is achieved, through which we propose that the nature of the inconsistency is not a violation to the essence of the QSL theory but an illusion caused by the finite precision in numerical simulations. We also present a generic method to overcome the inconsistent estimates and confirm its effectiveness in both amplitude-damping and phase-damping channels. Additionally, we show special cases which may restrict the QSL bound defined by "quantumness".

Ⅰ. INTRODUCTION

The maximal evolution speed of a quantum system is of basic importance in many fields of quantum physics, such as quantum metrology[1], quantum commutation and communication[2-4], quantum optimal control[5] and quantum thermodynamics[6]. One of the original bounds which predicts a minimal evolution time between two orthogonal pure states in a unitary process is $\tau_{QSL} = \pi\hbar/2\Delta E$, where $\tau_{QSL}$ denotes the QSL time and $\Delta E$ represents the variance of the energy of the initial state, was proposed by Mandelstam and Tamm(MT)[7]. A different bound in the form of $\tau_{QSL} = \pi\hbar/2E$, where the $\tau_{QSL}$ depends on the mean energy *E*, was derived by Margolus and Levitin(ML)[8] several decades later. By combining these two bounds, a final $\tau_{QSL} = \max\{\pi\hbar/2\Delta E, \pi\hbar/2E\}$ between two orthogonal pure states in a closed system was obtained. Since all systems are coupled to their environments essentially, finding the QSL time for an open system is highly desirable. To this end the MT type bound was first extended to a nonunitary process by Taddei *et al.*[9] and del Campo *et al.*[10]. Furthermore the ML type bound was extended to nonunitary processes by Deffner and Lutz[11]. Some other notions were also provided to generalize the QSL bound to an open system[12-16] and even to unify the metric of bounds[17,18]. Very recently Mirkin *et al.* [19] stated that some typical QSL bounds which are relevant to the definition of the average speed of evolution and depended explicitly on the actual evolution time are not close to the essence of the QSL theory for they give inconsistent estimates of the minimal evolution time in a damped JC model. We make a deep research on this puzzling inconformity and unveil the intrinsic reason of this phenomenon in this paper. Since the inconsistency always arises as soon as the limit of resolution of a calculation program is achieved, we present a generic method to eliminate the inconsistent estimates essentially. By using the modified bounds in both amplitude-damping and phase-damping channels, we confirm the effectiveness of our method. Additionally, special cases which may restrict the QSL bound defined by "quantumness" are observed.

*Corresponding author
 Email: 14171929@qq.com

The paper is structured as follows. In Sec. II we review three definitions of the QSL time which are used in this paper. In Sec. III we illustrate the inconsistent estimates which were presented by Mirkin *et al.* [19] and interpret this puzzling phenomenon in detail. Based on our interpretation, an effective method is proposed to solve the inconsistency in Sec. IV. Finally, we draw our conclusions in Sec. V.

II. THREE TYPICAL DEFINITIONS OF QSL TIME IN NONUNITARY PROCESSES

Three typical definitions of QSL bounds which we use in this paper are reviewed in this section. The first bound was originally presented in [9]. It comes from the fact that Bures angle leads to a lower bound for the Bures length. Namely,

$$B(\rho_0,\rho_t) = \arccos(F_B(\rho_0,\rho_t)) \leq \int_0^t \sqrt{\zeta_Q(t')/4}\, dt', \tag{1}$$

where $F_B(\rho_0,\rho_t) = \mathrm{Tr}(\sqrt{\sqrt{\rho_0}\rho_t\sqrt{\rho_0}})$ denotes the Bures fidelity[20] and $\zeta_Q(t') = \mathrm{Tr}[\rho(t)L^2(t)]$ [21] is the quantum Fisher information. Here the symmetric logarithmic derivative operator $L(t)$ is defined as $d\rho(t)/dt = [\rho(t)L(t)+L(t)\rho(t)]/2$. A QSL bound can be obtained immediately by simply transforming Eq. (1) as

$$t \geq \tau^{av} = \frac{B(\rho_0,\rho_t)}{v^{av}}, \tag{2}$$

where $v^{av} = (1/t)\int_0^t \sqrt{\zeta_Q(t')/4}\, dt'$ is defined as the "average speed of evolution".

The origin of the second bound which was presented in [11] is shown in Eq. (3),

$$\frac{dB(\rho_0,\rho_t)}{dt} \leq \left|\frac{dB(\rho_0,\rho_t)}{dt}\right|. \tag{3}$$

By using the von Neumann trace inequality and the Cauchy-Schwarz inequality respectively, Eqs. (4) and (5) are derived as follow:

$$2\cos(B)\sin(B)\dot{B} \leq \|L_t(\rho(t))\|_{op} \leq \|L_t(\rho(t))\|_{tr}, \tag{4}$$

$$2\cos(B)\sin(B)\dot{B} \leq \|L_t(\rho(t))\|_{hs}, \tag{5}$$

where $L_t$ is a positive generator which satisfies $\dot{\rho}(t) = L_t(\rho(t))$ and $\|\ \|_{op,hs,tr}$ denotes operator norm, Hilbert-Schmidt norm and trace norm, respectively. A unified QSL bound including MT and ML type bounds can be derived by combining Eqs. (4) and (5) after being

integrated over time:

$$t \geq \tau^{op,hs,tr} = \frac{\sin^2[B(\rho_0,\rho_t)]}{\min\{v^{op},v^{hs},v^{tr}\}}, \quad (6)$$

where $v^{op,hs,tr} = (1/t)\int_0^t \|L_t(\rho(t'))\|_{op,hs,tr} dt'$ is defined as the average velocity in frequency units.

The third bound is derived not by distance metric but the notion of the "quantumness"[22]. With the application of Cauchy-Schwarz inequality, a QSL bound is proposed as Eq. (7),

$$t \geq \tau^{quant} = \frac{\sqrt{Q(\rho_0,\rho_t)/2}}{v^{quant}}, \quad (7)$$

where $v^{quant} = (1/t)\int_0^t \|\rho_0,\dot{\rho}_{t'}\|_{hs} dt'$ is defined in a similar way as previous and $Q(\rho_0,\rho_t) = 2\|[\rho_0,\rho_t]\|_{hs}^2$ is presented to quantify the quantumness of a system[23,24].

III. THE INCONSISTENT ESTIMATES IN QUANTUM CHANNELS AND INTERPRETATION

In this section we focus on a puzzling phenomenon which has been presented by Mirkin *et al.* very recently[19] with using the three QSL bounds mentioned above to estimate the QSL time in quantum channels. The evolution of a given state in quantum channels can be expressed as follow,

$$\rho(t) = \sum_\mu K_\mu(t)\rho(0)K_\mu^+(t), \quad (8)$$

where the operators $K_\mu$ are the so-called Kraus operators[25] and satisfy $\sum_\mu K_\mu^+ K_\mu = 1$ for all $t$.

First we make a calculation in an amplitude-damping channel which represents the dissipative interaction between a qubit and its environment. The Hamiltonian model for the process is[26,27]

$$H_{AD} = \omega_0 \sigma_+ \sigma_- + \sum_k \omega_k a_k^+ a_k + (\sigma_+ B + \sigma_- B^+), \quad (9)$$

where $B = \sum_k g_k a_k$ with $g_k$ being the coupling constant. $\omega_0$ is the transition frequency of the qubit, and $\sigma_\pm$ denote the raising and lowering operators related to the qubit. Here the index $k$ labels the reservoir field modes with frequencies $\omega_k$, and $a_k(a_k^+)$ is their annihilation(creation) operator. We use a damped JC model with

$$J(\omega) = \frac{\gamma_0 \lambda^2}{2\pi[(\omega_0+\Delta-\omega)^2 + \lambda^2]}, \quad (10)$$

where $\lambda$ defines the spectral width and $\gamma_0$ quantifies the coupling strength, $\Delta$ is a detuning between the peak frequency of the spectral density and $\omega_0$. In our case $\Delta$ is considered to be zero to simplify the calculation. Then the decoherence function is

$$G(t) = e^{-\lambda t/2}[\cosh(\frac{dt}{2}) + \frac{\lambda}{d}\sinh(\frac{dt}{2})], \tag{11}$$

where $d = \sqrt{\lambda^2 - 2\gamma_0 \lambda}$. The dynamics is a Markovian process where $\gamma_0 < \lambda/2$, otherwise it is a non-Markovian one[26,27]. In this model the Kraus operators are given by

$$K_1(t) = \begin{pmatrix} 1 & 0 \\ 0 & G(t) \end{pmatrix}, \quad K_2(t) = \begin{pmatrix} 0 & \sqrt{1-G^2(t)} \\ 0 & 0 \end{pmatrix}. \tag{12}$$

For an arbitrarily initial state, the asymptotic final stationary state in this channel is always to be the ground state $|0\rangle$. We use $|\varphi_0\rangle = \frac{\sqrt{2}}{2}(|0\rangle + |1\rangle)$ as our initial state. The decoherence function and the trace distance between the evolved state $\rho_t$ and its finial stationary state $\rho_{stat}$ in Markovian and non-Markovian processes are shown in Fig. 1 (a) and (b), respectively. Through the evolution of the decoherence function (or the trace distance), we find that the finial state of the channel is approached at a finite time, i.e. the QSL time between the initial state and the finial state is limited. On the other hand, when we use all the three QSL bounds to estimate the QSL time as a function of the actual evolution time $t$ both in Markovian and non-Markovian environments, which are respectively shown in Fig. 1 (c) and (d), we find that the bounds grow indefinitely even after the time at which the finial state is achieved in Fig. 1 (a) ((b)). Intuitively, this phenomenon leads to inconsistent estimates and Mirkin *et al.* have interpreted these inconsistent estimates as the dependence of the actual evolution time. Therefore they argued that these bounds are not conforming to the essence of the QSL theory.

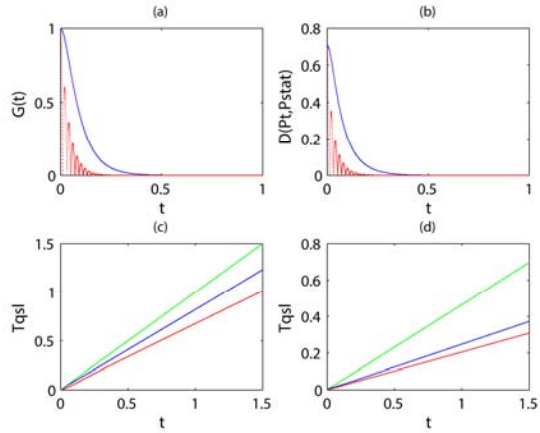

Fig. 1. (a) The decoherence function and (b) the trace distance between the evolved state and its finial stationary state in an amplitude-damping channel. $\tau_{qsl}^{av}$ (blue line), $\tau_{qsl}^{op}$ (red line), $\tau_{qsl}^{quant}$ (green line) in (c) a Markovian environment with $\gamma_0/\lambda = 0.4$ and (d) a non-Markovian

environment with $\gamma_0/\lambda=20$. The blue curve in (a) ((b)) means the same Markovian environment and the red one means the same non-Markovian environment.

Then we investigate what is the case in a phase-damping channel which describes a pure dephasing type of interaction between a qubit and a bosonic reservoir. The Hamiltonian of the total system is written as[28]

$$H_{PD} = \omega_0 \sigma_z + \sum_k \omega_k a_k^+ a_k + \sum_k \sigma_z (g_k a_k + g_k^* a_k^+). \tag{13}$$

The spectral density of the environmental modes is supposed to be Ohmic-like:

$$J(\omega) = \frac{\omega^s}{\omega_c^{s-1}} e^{-\omega/\omega_c}, \tag{14}$$

where $\omega_c$ is the reservoir cutoff frequency. By changing the parameter $s$, the spectral density is named sub-Ohmic($s<1$), Ohmic($s=1$), and super-Ohmic($s>1$) respectively. Here the Kraus operators are given by

$$K_1(t) = \begin{pmatrix} 1 & 0 \\ 0 & r(t) \end{pmatrix}, \quad K_2(t) = \begin{pmatrix} 0 & 0 \\ 0 & \sqrt{1-r^2(t)} \end{pmatrix}, \tag{15}$$

where the decoherence function $r(t)$ reads

$$r(t) = \exp[-\int_0^t \gamma(t') dt'], \tag{16}$$

and the dephasing rate $\gamma(t)$ is

$$\gamma(t) = \omega_c [1+(w_c t)^2]^{-s/2} \Gamma(s) \sin[s \arctan(\omega_c t)], \tag{17}$$

with $\Gamma(s)$ being the Euler function. By using the same initial state and comparing the decoherence function (or the trace distance) with the evolution of the QSL time, which are respectively shown in Fig. 2 (a) (or (b)) and (c) ((d)), similar inconsistent estimates are appeared in this channel. It is noteworthy that the QSL time which is calculated by using the third QSL bound with our initial state is always zero in this channel and we will give a discussion on this special case in next section.

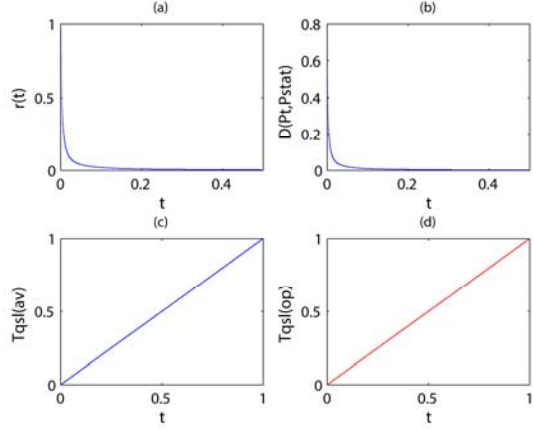

Fig. 2. (a) The decoherence function and (b) the trace distance between the evolved state and its finial stationary state in a phase-damping channel (s=1). The evolution of (c) $\tau_{qsl}^{av}$ and (d) $\tau_{qsl}^{op}$ in the same channel.

Now let us deeply consider these inconsistent estimates. Do they really imply that the three bounds mentioned above go against the essence of the QSL theory? As we have known, an arbitrary initial state will eventually evolve to a stationary state in any quantum channels where the actual evolution time between the initial state and the final stationary state is infinite. Namely, an initial state should only stop evolving in the channel at $t \to \infty$, through which it means that the final stationary state is reached. It leads to the fact that the QSL time between an initial state and the final stationary state of a quantum channel remains unchanged only at $t \to \infty$, which is corresponding to the dynamics of the QSL time shown in Fig 1 and Fig 2. Meanwhile, the decoherence function (or the trace distance) should also reach zero at $t \to \infty$. However, the precision of a calculation program which is used to calculate the parameters is finite. That is to say, the decoherence function (or the trace distance) will be indistinguishable from zero at a finite time $\tau^{cri}$ at which the limit of resolution of the calculation program is achieved. Therefore the final stationary state which is considered to be reached at a finite time by using the trace distance in [19] is not the true final stationary state but an approximation which is indistinguishable to the exact final stationary state due to the finite precision. This exactly causes the inconsistent estimates that the final state is reached at a finite time but the QSL bound grows indefinitely. Accordingly, we deem that the intrinsic reason of inconsistent estimates is not a departure of the essence of the QSL theory, but an illusion caused by the finite precision of any calculation programs. It is clearly illustrated in Fig. 3.

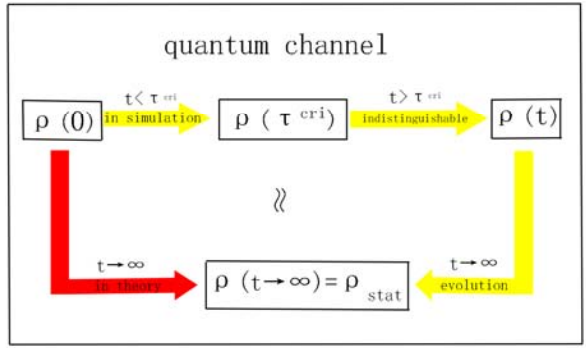

Fig. 3. The intrinsic reason which causes the inconsistent estimates.

Ⅳ. MODIFICATION FOR ELIMINATING THE INCONSISTENCY

According to the cause of the inconsistent estimates, as shown explicitly in Fig. 3, we present a generic method to essentially solve the problem. That is using $\tau^{cri}$, which denotes the maximally actual evolution time in simulation for the limit of resolution of the calculation program is achieved, to replace the actual evolution time $t$ once $t$ is beyond $\tau^{cri}$ in any numerical calculations. This substitution can be easily implemented and proved in math. The three bounds mentioned above are used again as paradigms.

For the first bound, when the actual evolution time $t$ is less than $\tau^{cri}$, the state will keep evolving and the inconsistent estimate should not emerge. So the derivation of the bound does not need any alternation. It is not the case once the actual evolution time $t$ reaches $\tau^{cri}$ after which the evolution of the state supposes to stop and $\rho(\tau^{cri}) \approx \rho_{stat}$ remains unchanged. Then Eq. (1) can be rewritten in the form of

$$B(\rho_0,\rho_t) = \arccos(F_B(\rho_0,\rho_t)) \leq \int_0^{\tau^{cri}} \sqrt{\zeta_Q(t')/4}dt' + \int_{\tau^{cri}}^{t} \sqrt{\zeta_Q(t')/4}dt'. \quad (18)$$

For the left hand of the equation, since the state is not changed any more, i.e., $\rho(\tau^{cri}) = \rho(t) = \rho_{stat}$, we get $B(\rho_0,\rho_t) = \arccos(F_B(\rho_0,\rho_t)) = \arccos(F_B(\rho_0,\rho_{\tau^{cri}})) = B(\rho_0,\rho_{\tau^{cri}})$. The second item in the right hand of the equation is zero for a similar reason. Eq. (18) is reduced to form (19),

$$B(\rho_0,\rho_{\tau^{cri}}) \leq \int_0^{\tau^{cri}} \sqrt{\zeta_Q(t')/4}dt'. \quad (19)$$

An alternative QSL time for $t \geq \tau^{cri}$ is then obtained immediately,

$$\tau_{qsl}^{av} = \frac{B(\rho_0,\rho_{\tau^{cri}})}{v^{avc}}, \quad t \geq \tau^{cri}, \quad (20)$$

where $v^{avc} = (1/\tau^{cri})\int_0^{\tau^{cri}} \sqrt{\zeta_Q(t')/4} dt'$. The ultimate QSL time in actual calculations is shown as Eq. (21),

$$\tau_{qsl}^{av} = \begin{cases} \dfrac{B(\rho_0,\rho_t)}{v^{av}}, & t < \tau^{cri} \\ \dfrac{B(\rho_0,\rho_{\tau^{cri}})}{v^{avc}}, & t \geq \tau^{cri} \end{cases}. \tag{21}$$

For the second bound, we use $\tau^{op}$ which had been proved to provide the sharpest bound on the QSL time in [11] as example. Integrating Eq. (4) over time we get

$$\sin^2[B(\rho_0,\rho_t)] \leq \int_0^t \|L_t(\rho(t'))\|_{op} dt'. \tag{22}$$

There is still no need to modify the bound if $t < \tau^{cri}$ as the illusion is not broken out yet. Otherwise we rewrite Eq. (22) in the form of

$$\sin^2[B(\rho_0,\rho_t)] \leq \int_0^{\tau^{cri}} \|L_t(\rho(t'))\|_{op} dt' + \int_{\tau^{cri}}^t \|L_t(\rho(t'))\|_{op} dt'. \tag{23}$$

Since the state stops evolving at $\tau^{cri}$ in numerical simulations, $\sin^2[B(\rho_0,\rho_t)] = \sin^2[B(\rho_0,\rho_{\tau^{cri}})]$ and $\int_{\tau^{cri}}^t \|L_t(\rho(t'))\|_{op} dt' = 0$ hold once again and a fixed bound is

$$\tau_{qsl}^{op} = \dfrac{\sin^2[B(\rho_0,\rho_{\tau^{cri}})]}{v^{opc}}, \quad t \geq \tau^{cri}, \tag{24}$$

where $v^{opc} = (1/\tau^{cri})\int_0^{\tau^{cri}} \|L_t(\rho(t'))\|_{op} dt'$. Therefore a total QSL time is

$$\tau_{qsl}^{op} = \begin{cases} \dfrac{\sin^2[B(\rho_0,\rho_t)]}{v^{op}}, & t < \tau^{cri} \\ \dfrac{\sin^2[B(\rho_0,\rho_{\tau^{cri}})]}{v^{opc}}, & t \geq \tau^{cri} \end{cases}. \tag{25}$$

With a similar analysis on the derivation of the third bound, the QSL time defined by "quantumness" of the system in numerical simulations is arrived at

$$\tau_{qsl}^{quant} = \begin{cases} \dfrac{\sqrt{Q(\rho_0,\rho_t)/2}}{v^{quant}}, & t < \tau^{cri} \\ \dfrac{\sqrt{Q(\rho_0,\rho_t)/2}}{v^{quantc}}, & t \geq \tau^{cri} \end{cases}, \tag{26}$$

where $v^{quantc} = (1/\tau^{cri})\int_0^{\tau^{cri}} \|\rho_0,\dot\rho_{t'}\|_{hs} dt'$.

Now we confirm the effectiveness of our method in both amplitude-damping and phase-damping channels which are used in previous section. A most important thing in our method is to calculate $\tau^{cri}$. It is intuitive that $\tau^{cri}$ depends on the limit of resolution of a calculation program. With a same calculation program, many time-dependent functions which gradually approach to zero can be used to calculate $\tau^{cri}$. Such as the decoherence function and the trace distance between the evolved state and the final stationary state of the quantum channel, which are mentioned in previous part. Obviously higher limit of resolution leads to better accuracy but longer $\tau^{cri}$. For our aim is to check if the modified bounds have fixed the inconsistent estimates, we set the limit of resolution of the calculation program at $10^{-6}$ to reduce the calculation time.

The evolutions of the three modified QSL bounds in Markovian and non-Markovian processes in an amplitude-damping channel are shown in Figs. 4 (a) and (b), respectively. It can be seen that all the three QSL time keep increasing until $\tau^{cri}$ is reached and then remain unchanged. These sufficiently indicate that the inconsistent estimates are eliminated. By comparing Fig. 4 (a) and (b), it is obviously seen that both the QSL time and the maximally actual evolution time $\tau^{cri}$ in a non-Markovian environment are shorter than those in a Markovian environment. It therefore indicates that non-Markovianity can speed up quantum evolution. Since the QSL bounds used here are saturated whenever the equality in Eqs. (2), (6) and (7) are achieved, i.e. $t = Tqsl$ is obtained, the tightness of the three modified bounds can be expressed by $Tqsl/t$. As shown in Fig. 4 (c) and (d), the third bound which is defined through "quantumness" is the tightest one in both Markovian and non-Markovian environments.

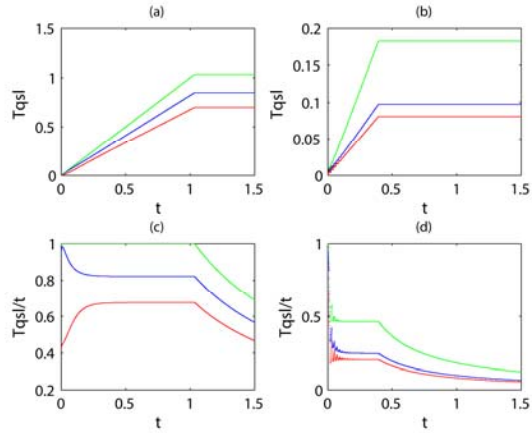

Fig. 4. The three modified QSL time in (a) a Markovian environment with $\gamma_0/\lambda=0.4$ and (b) a non-Markovian environment with $\gamma_0/\lambda=20$ in an amplitude-damping channel. The tightness of the three bounds in (c) a Markovian environment and (d) a non-Markovian environment with the same parameters. The blue line represents $\tau_{qsl}^{av}$; the red line, $\tau_{qsl}^{op}$; the green line, $\tau_{qsl}^{quant}$.

The first two modified QSL bounds and their tightness in a phase-damping channel are shown in Fig. 5. It can be seen that the inconsistent estimates are also disappeared in this case. The two bounds are both saturated until $\tau^{cri}$ is reached. As we have mentioned in previous section, the third bound is always zero for the "quantumness" which is defined in section Ⅱ is always zero with our initial state and its evolved state in this channel. Since our given initial state should evolve in this channel, it is not in accordance with the fact. A similar scenario also arises in an amplitude-damping channel if we use the excited state $|1\rangle$ as an initial state (not illustrated). Accordingly, the applicability of the third bound, which uses the "quantumness" as a metric of the QSL bound, may need a further discussion.

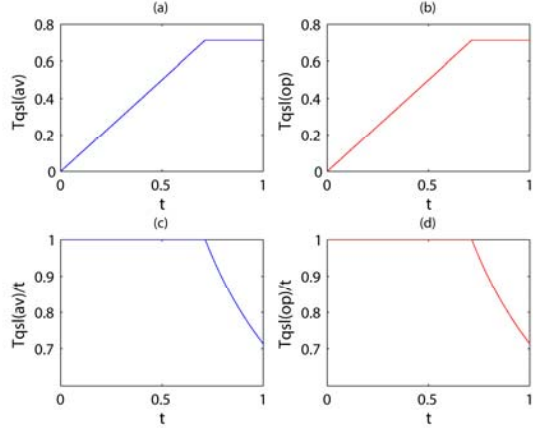

Fig. 5. (a) $\tau_{qsl}^{av}$, (b) $\tau_{qsl}^{op}$ and (c), (d) their tightness in a phase-damping channel with s=1.

Ⅴ. CONCLUSIONS

The maximal speed of evolution of a quantum system which is relevant to the minimal evolution time between two distinguishable states is a fundamental limit for the application of quantum physics. Many QSL bounds have been presented in open systems. Mirkin *et al*. claim that some typical QSL bounds, which depend on the actual evolution time, are not corresponding to the essence of the QSL theory for they grow infinitely but the final state is reached finitely. We make a deep study on this inconsistency and find what causes a finial state be reached finitely is that the limit of resolution of the calculation program is achieved. Therefore we deem that the inconsistent estimates are only an illusion caused by the finite precision of a calculation program in any numerical simulations. Based on the cause of the puzzling phenomenon, we offer a generic method to conquer the problem. The validity of our method is verified in both amplitude-damping and phase-damping channels. Moreover, we point out special cases in which the bound defined by

"quantumness" is inapplicable.

ACKNOWLEDGMENTS

C Wang acknowledges the support from EU FP7 Marie-Curie Career Integration Grant under Grant 631883, and the Royal Society Research Grants under Grant RG150036. X Liu and W Wu acknowledge the support from the Fundamental Research Funds for the Central Universities under the project number 2015IA006.